\documentclass[a4paper]{article}
\usepackage{xcolor}
\usepackage{multirow}
\usepackage{amsmath}
\usepackage{graphicx}
\usepackage{setspace}
\usepackage{mdwlist}
\usepackage{enumitem}

\usepackage{tikz}
\usetikzlibrary{shapes.geometric, arrows}
\tikzstyle{stepblock} = [rectangle, rounded corners, minimum width=8cm, minimum height=1.2cm, text centered, text width=8cm, draw=black]
\tikzstyle{arrow} = [thick,->,>=stealth]

\DeclareMathOperator{\Tr}{Tr}

\usepackage{INTERSPEECH2020}

\title{Variable frame rate-based data augmentation to handle speaking-style variability for automatic speaker verification}
\name{Amber Afshan$^1$, Jinxi Guo$^1$ \sthanks{\hspace{0.35em} These authors have equal contribution}, Soo Jin Park$^1$\hspace{-0.05em}\footnotemark[1], Vijay Ravi$^1$\hspace{-0.05em}\footnotemark[1], Alan McCree$^2$, and Abeer Alwan$^1$}
\address{
  $^1$Dept. of Electrical and Computer Engineering, University of California, Los Angeles, USA\\
  $^2$Human Language Technology Center of Excellence, Johns Hopkins University, Baltimore, USA}
\email{amberafshan@g.ucla.edu}

\begin{document}

\maketitle
\begin{abstract}
The effects of speaking-style variability on automatic speaker verification were investigated using the UCLA Speaker Variability database which comprises multiple speaking styles per speaker. An x-vector/PLDA (probabilistic linear discriminant analysis) system was trained with the SRE and Switchboard databases with standard augmentation techniques and evaluated with utterances from the UCLA database. The equal error rate (EER) was low when enrollment and test utterances were of the same style (e.g., 0.98\% and 0.57\% for read and conversational speech, respectively), but it increased substantially when styles were mismatched between enrollment and test utterances. For instance, when enrolled with conversation utterances, the EER increased to 3.03\%, 2.96\% and 22.12\% when tested on read, narrative, and pet-directed speech, respectively. To reduce the effect of style mismatch, we propose an entropy-based variable frame rate technique to artificially generate style-normalized representations for PLDA adaptation. The proposed system significantly improved performance. In the aforementioned conditions, the EERs improved to 2.69\% (conversation -- read), 2.27\% (conversation -- narrative), and 18.75\% (pet-directed -- read). Overall, the proposed technique performed comparably to multi-style PLDA adaptation without the need for training data in different speaking styles per speaker.  
\end{abstract}

\section{\label{sec:intro}Introduction}
An individual often varies his/her speaking style in day-to-day situations.  Reading aloud, having a conversation, and talking to animals result in different acoustic properties in the speech signal. For instance, acoustical differences between read and conversational speech include different speaking rates and inconsistent pauses between words. There are also variations in the number and type of phonological phenomena observed.  For example, vowels are modified or reduced in conversational speech, and word-final plosive bursts are not released while it is not the case in read speech~\cite{picheny_speaking_1986}. Similar differences are observed across other speaking styles as well~\cite{eskenazi_trends_1993}. 

When the acoustic properties of an individual's speech differ between the enrollment and test utterances, automatic speaker verification (ASV) system performance generally degrades~\cite{shum_unsupervised_2014}. There are two categories of within-speaker variability that causes  such difference: extrinsic and intrinsic variability. \textit{Extrinsic variability} is associated with factors not directly related to the speaker's behavior (e.g., recording conditions, channel types, and environmental noise). There has  been  considerable progress in studying the effects of extrinsic variability on ASV performance~\cite{snyder_x-vectors_2018, guo_robust_2017, guo_deep_2018, guo_cnn-based_2017, guo_speaker_2016, sarangi_improved_2020}. On the other hand, \textit{intrinsic variability} is related to the speaker's conscious and/or unconscious behavior that can influence speech signal production. Studies showed that ASV performance degraded due to intrinsic variabilities--vocal effort, speaking styles, speaking rate, loudness, emotional state and physical status~\cite{shriberg_effects_2008, shriberg_does_2009, chen_compensation_2012}.

Speaking style variability is a type of intrinsic variability which can make acoustic characteristics considerably different within a speaker. However, only a limited number of studies have investigated the effects of style variability on ASV performance.  \textit{Style factors} are shown to be present in widely used speaker representations~\cite{williams_disentangling_2019} such as i-vectors~\cite{dehak_front-end_2011} and x-vectors~\cite{snyder_x-vectors_2018}.
ASV performance degradation due to style mismatch between the enrollment and test utterances were systematically analyzed in~\cite{park_speaker_2016, park_using_2017, park_towards_2018}. To alleviate the degradation due to style variabilities, some studies proposed the use of a joint factor analysis framework \cite{shriberg_does_2009, chen_compensation_2012}.  In~\cite{zhang_analysis_2018}, curriculum-learning based transfer learning was done using neutral/physical stressed as well as read and spontaneous speech to compensate for style mismatches during testing. Note that the compensation techniques proposed in these studies require a variety of speaking styles per speaker to train the systems, i.e., the training data includes all the styles occurring in the test utterances~\cite{zhang_analysis_2018}. However, one might not always have prior knowledge of the speaking style of the test utterances. 

One can expect that including various speaking styles in the training data may improve the speaking-style robustness of the system. However, corpora with sufficient numbers of speakers speaking with different styles are not available. A widely-used approach to address insufficient data to train different conditions in ASV is \textit{data augmentation} using artificially generated data. Augmentation strategies include adding variations of noise, reverberation~\cite{heigold_end--end_2016,snyder_x-vectors_2018}, collecting additional domain-specific data~\cite{zhang_analysis_2018}, and synthesizing data~\cite{rituerto-gonzalez_data_2019}. 
 Yet, for style variability, artificially synthesizing speaking styles is not yet reliable enough to be applied~\cite{wang_uncovering_2017, williams_disentangling_2019}. In this work, we propose the use of a variable frame rate (VFR) approach to generate style-normalized representations to perform data augmentation.

The rest of the paper is organized as follows. Section~\ref{sec:databases} describes the databases used. The proposed approach is detailed in Section~\ref{sec:system}.  
Section~\ref{sec:expnres} provides the experimental setup and discusses the results, and we conclude with Section~\ref{sec:conclusion}.

\section{\label{sec:databases}Databases}

\subsection{\label{sec:uclasvd}The UCLA Speaker Variability Database}

In order to systematically study both within- and between-speaker variability, a multi-speaker speech database including multiple speech tasks per speaker is needed. The UCLA Speaker Variability Database~\cite{kreiman_relationship_2015, keating_new_2019} provides multiple recordings of speakers in a variety of speech tasks and on multiple occasions. Audio recordings were done in a sound-attenuated booth with a sampling rate of 22 kHz. 

Speech tasks from the database used for this study include reading sentences to represent scripted speaking style ($\approx 75$ sec); narrating a recent neutral, happy, or annoying conversation to represent unscripted affective speech ($\approx$ 30 sec each); making a telephone call to a familiar person to represent unscripted conversational style (60--120 sec); and talking aloud to pets in a video, providing pet-directed speech, which typically has exaggerated prosody (60--120 sec). 
 
\subsection{\label{sec:nist}Databases for Training the ASV System}

The Speaker Recognition Evaluation (SRE) databases developed by NIST are often used to train ASV systems.
We used the NIST SRE 04, 05, 06, 08 and 10 databases~\cite{przybocki_nist_2004,przybocki_nist_2006,martin_nist_2009} along with the Switchboard  II  corpus, phase 2~\cite{graff_switchboard-2_1999} for this purpose. 
The sampling rate for these databases is 8~kHz. 

Note that although the SRE and Switchboard databases offer many recordings from a large number of speakers with multiple speech tasks, they do not provide multiple speech tasks per speaker under controlled recording environments. Additionally, they do not provide metadata regarding speaking style. Thus, the UCLA Speaker Variability Database is more suitable for detailed analyses of the effects of style variability. The UCLA dataset was downsampled to match the sampling rate of the training databases.

\section{\label{sec:system}Proposed Approach}
\subsection{\label{sec:baseline}Automatic Speaker Verification System}
The Kaldi~\cite{povey_kaldi_2011} SRE16 recipe was used to develop a x-vector/PLDA ASV system~\cite{snyder_x-vectors_2018}. The input acoustic features were 23-dimensional mel-frequency cepstral coefficients (MFCCs) with a frame length of 25~ms and a frame shift of 10~ms, which were mean normalized over a sliding window of up to 3~secs. 
Standard extrinsic data augmentation (as in the recipe) was applied on the training-data lists of both x-vector and PLDA.

A widely-used strategy to attenuate within-speaker variability is to train the PLDA with data for the conditions of variability from each speaker~\cite{garcia-romero_unsupervised_2014, garcia-romero_multicondition_2012}.
Although this strategy has been mainly used for external sources of variability (e.g, noise, channel, etc.)~\cite{ snyder_x-vectors_2018, garcia-romero_multicondition_2012}, it could be also applied to deal with the speaking style variability.
However, sufficient amount of data is not available in the UCLA database to train a robust PLDA in this manner. Therefore, a PLDA model was trained with the previously mentioned training list and the in-domain adaptation (using the version provided in Kaldi) was performed with the UCLA database. 
The experimental configurations for adaptation will be described in Section~\ref{sssec:plda}.

\subsection{\label{sec:vfr}Data Augmentation using Variable Frame Rate}


In cases when multiple speaking styles per talker are not available in the training dataset, 
a method to artificially generate speaking style-normalized variants for augmentation is required.
We propose to use the entropy-based variable frame rate to generate such variants. 
As mentioned earlier, some of the key differences across speaking styles are speech rate, long pauses, changes in the duration of individual sounds, boundary articulation. In this work, we aimed at reducing the effects of such acoustic differences on ASV performance. Specifically, we propose to generate style-normalized speaker representations by applying the entropy-based variable frame rate approach~\cite{you_entropy-based_2004}. 

\subsubsection{\label{sec:entropy}Entropy Computation}
Consider a random variable $\nu \in \mathcal{R}^K$ where $p(\nu)$, the probability distribution function (PDF) of $\nu$ is a $K$-dimensional Gaussian. Let $\mu$ and $\Sigma$ be the mean and covariance matrix of the random variable. The entropy can be calculated as:

\begin{equation}
\small
 \begin{split}
    H(\nu) &= - \int p(\nu) \ln{p(\nu)} d{\nu} \\
    &= - \int p(\nu) \Bigg[ -\frac{1}{2}(\nu-\mu)^T \Sigma^{-1} (\nu-\mu) - \ln{|2\pi\Sigma|^{ \frac{1}{2}}}\Bigg] d{\nu}\\
     &= \frac{K}{2} + \frac{1}{2} \ln{|2\pi \Sigma|}
\end{split}   
\end{equation}

To facilitate faster computation and to avoid an ill-posed problem when the random variable's covariance matrix is not full rank~\cite{you_entropy-based_2004}, the following approximation is used to calculate the entropy:
\begin{equation}
    H(\nu) \approx K \ln{\sqrt{2\pi}} +  \ln{\Tr{\Sigma}}
\end{equation}

\subsubsection{\label{vfrcomp}Implementation}

\begin{figure}[h]
    \centering
     \begin{tikzpicture}[node distance=1.5cm]
        \node (step0) [stepblock] {Hamming window of 25~ms and 2.5~ms frame shift (oversampled)};
        \node (step01) [stepblock, below of=step0] {Extract mel-filter spectrum};
        \node (step1) [stepblock, below of=step01] {Initialize buffer for 30~ms mel-filter spectrum sequence};
        \node (step2) [stepblock, below of=step1] {Compute the entropy curve every 15~ms};
        \node (step3) [stepblock, below of=step2] {Set frame picking thresholds $T_1-T_3$ using the signal statistics};
        \node (step4) [stepblock, below of=step3, yshift=-0.15cm] {Pick frame rate based on $ H(\nu_i)$ as: \\ 5, 7.5~ms (more samples), 10~ms (no change) and 12.5~ms (less samples)};
        \draw [arrow] (step0) -- (step01);
        \draw [arrow] (step01) -- (step1);
        \draw [arrow] (step1) -- (step2);
        \draw [arrow] (step2) -- (step3);
        \draw [arrow] (step3) -- (step4);
     \end{tikzpicture}
     \caption{Overview of the entropy-based variable frame rate approach.}
    \label{fig:vfr_bd}
\end{figure}
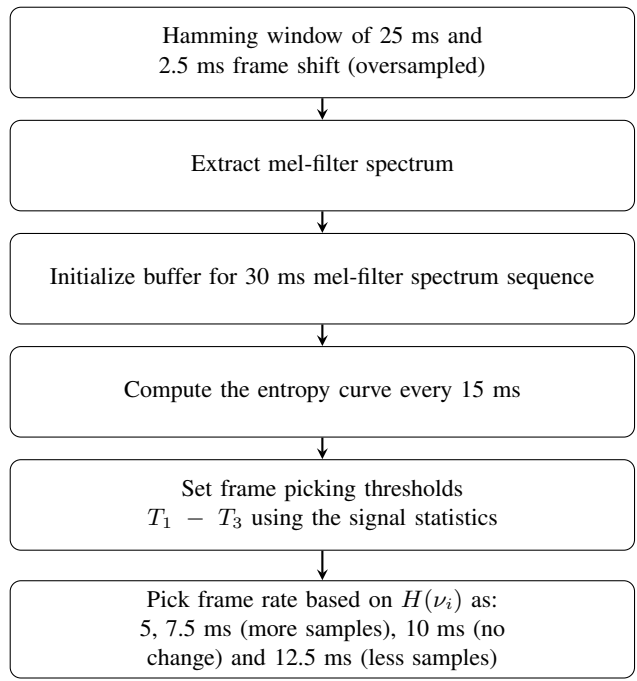

The variable frame rate approach dynamically changes the frame rate based on inter-frame entropy using the steps shown in Figure~\ref{fig:vfr_bd}. First, a signal is windowed using 25~ms Hamming window by first sampling with frame shift of 2.5~ms, a much lower value than widely-used 10~ms frame shift. With these densely sampled, or ``oversampled'' frames, varying frame rate becomes a simple task of retaining frames selectively. Mel-filter spectra are then computed. The frames spanning a duration of 30~ms are then used to calculate the entropy curve using the local entropy every 15~ms.  
VFR was carried out by comparing the signal's entropy to certain thresholds in order to calculate the frame picking rate in the extraction of MFCCs. Using the entropy curve of the speech signal $H(\nu_i)$, $i = 1,..., N$, the frame-picking thresholds $T_1, T_2, T_3$ are set as in Equation~\ref{eq:thresholds}.

\begin{equation}
\label{eq:thresholds}
    \begin{cases}
    T_1 &= \omega_1 M_{max} + (1 - \omega_1) M_{med}\\
    T_2 &= (1 - \omega_2)M_{max} + \omega_2M_{med}\\
    T_3 &= (1 - \omega_3)M_{med} + \omega_3M_{min},\\
    \end{cases}
\end{equation}
where $\omega_1$, $\omega_2,$ and $\omega_3$ are weighting parameters of values 0.7, 0.8, and 0.5, respectively. $M_{max}$, $M_{med},$ and $M_{min},$ are the maximum, median, and minimum of the entropy curve, respectively. In this implementation, the x-vector extractor is trained using a frame shift of 10~ms. Hence, frame rates of 5~ms ($ H(\nu_i) \geq T_1$)  and 7.5~ms ($ T_1 > H(\nu_i) \geq T_2$) are used to  obtain more frames from the regions where the signal has rapid changes of information. A 10~ms frame shift is used when entropy is close to average ($T_2 > H(\nu_i) \geq T_3$). Whereas the frame rate is 12.5~ms ($ T_3 > H(\nu_i)$) when the signal has low information gain, so that we obtain lesser frames from the region.  

Fast speech rate and/or short pause can lead to a rapid change of information in spectral characteristics between frames resulting in a higher inter-frame entropy. Style variability may also cause a decrease in speech rate that could result in a low inter-frame entropy. 
That is, speaking style variability is at least partially reflected in the inter-frame entropy.
Thus, extracting features maintaining consistent inter-frame entropy as much as possible could, in-turn, normalize the effects of style variability such as speech rate and duration of individual sounds. 

Based on the above assumption, VFR was used to generate partially style-normalized utterance representations. This approach is expected to be more robust than varying the speaking rate of the entire utterance because the variations within an utterance and within speaking style are not always uniform due to speaker characteristics, context of the conversation, emotion, etc.


\section{\label{sec:expnres}Experiments and Results}
\subsection{\label{sec:exp}Experimental Setup}
\subsubsection{\label{sec:datastats} Database Statistics}
A randomly selected subset of 50 female and 50 male speakers from the UCLA database was set aside as the ``development set''. The remaining subset of 50 female and 50 male speakers was used as the ``evaluation set''. The evaluation set was further split into ``enrollment'' and ``test'' set. 

In order to analyze the effect of style variability on system performance, the effect of phonetic variability across utterances needs to be negligible. Based on the reports that 30-sec utterances cover enough phonetic variability to capture speaker-specific information~\cite{hasan_duration_2013}, 30-sec long speech samples were used both for enrollment and test utterances. Table~\ref{tab:stats} shows the number of speech samples from the UCLA database used in this experiment. Note that at least 1 min of speech is required per speaker to generate style-matched enrollment -- test utterance pairs. Because the majority of speakers in the UCLA database did not have enough speech in the narrative and pet-directed speaking styles, style-matched conditions for those styles were omitted. This resulted in 14 different evaluation combinations. All possible trials were generated for all the styles, which resulted in more non-target trials than target trials. 

\begin{table}[h]
    \centering
        \caption{Number of utterances in distributed across each set for the UCLA database}
    \label{tab:stats}
\resizebox{\linewidth}{!}{%
\begin{tabular}{c|cccc}
\toprule
\textbf{Speaking Style} & \textbf{read} & \textbf{narrative} & \textbf{conversation} & \textbf{pet-directed} \\ \midrule
Development set & 196 & 36 & 184 & 19 \\
Enrollment set  & 102 & 35 & 99  & 16 \\ 
Test set        & 101 & 35 & 88  & 16 \\ \bottomrule
\end{tabular}%
}
\end{table}

\subsubsection{\label{sssec:plda}PLDA Adaptation Configurations}


The PLDA trained on SRE and Switchboard data is adapted using the development set from the UCLA database.
Recall that the major focus in this paper is data augmentation using VFR for PLDA adaptation. Hence, we designed the below five different adaptation configurations to experimentally analyze the advantages of the proposed technique:  

\begin{itemize}[label={}, leftmargin=*]

    \item \textbf{Baseline:} In-domain data with a single speaking style, the same as that of the enrollment set, is used (development set size $X$).

    \item \textbf{Extrinsic augmentation:} Extrinsic variability is added using artificial data augmentation (development set size $5X$). The implementation here is similar to the one in x-vector training~\cite{snyder_x-vectors_2018}, but we use all the extrinsic variants and not a subset. We add music, noise and babble from the MUSAN corpus~\cite{snyder_musan_2015} and reverb by convolving with simulated room impulse responses~\cite{ko_study_2017}.  

    \item \textbf{VFR normalization:} Entropy-based VFR normalization is applied to the development data of a single speaking style (development set size $X$). This generates partially style-normalized development set.
    
    \item \textbf{[Proposed] VFR normalization augmentation:} Both the original representations of the development data and their style-normalized counterparts, obtained by performing VFR, were used (development set size $2X$). 
    
    \item \textbf{Multi-style:} Multiple speaking styles from the in-domain data were used (development set size $4X $). 

\end{itemize}

In the baseline, extrinsic augmentation, VFR normalization, and VFR normalization augmentation configurations, the speaking style used in the development set matched that of the enrollment utterances. For instance, when enrolling with \textit{read} and testing with other styles, the development set for PLDA adaptation contained only \textit{read} sentences. In contrast, all styles in the development set were used in the multi-style configuration.

The baseline configuration was used to assess the effects of speaking style variability on ASV performance, as well as to establish baseline performance to be compared with the other configurations. 
The extrinsic augmentation configuration represents standard techniques that increase the amount of data, and it was used to understand how the proposed VFR data augmentation does when compared to it. 
The VFR normalization configuration was used to analyze the effectiveness of style-normalization with the VFR approach and also to assess if style-normalization alone would be enough to compensate for style variability. 

 Note that the multi-style configuration is the best-case scenario, but it is not realistic to assume that one can obtain all speaking styles for each speaker. 

\subsection{\label{sec:rd} Results and Discussion}
System performance in terms of the EER for the UCLA database is shown in Table~\ref{tab:ucla}. Statistical significance was verified using McNemar's test~\cite{mcnemar_note_1947}. Unless mentioned explicitly, all performance differences reported in this section are significant with $p<0.005$.

\begin{table}[h]
\caption{Performance in terms of EER (\%) on the UCLA database.} 
\label{tab:ucla}
\resizebox{\linewidth}{!}{
\begin{tabular}{ c  r | r r r r }
\toprule
  & & \multicolumn{4}{c}{\textbf{Test}} \\ \cline{3-6} 
  & \textbf{Enroll} & \textbf{read}  & \textbf{narrative} &  \textbf{conversation} & \textbf{pet-directed} \\  \midrule 
 \parbox[t]{4mm}{\multirow{4}{*}{\rotatebox[origin=c]{90}{\parbox[c]{1.5cm}{\centering \textbf{Baseline}}}}} &
  \textbf{read}         & 0.98   & 2.20  & 2.25  & 15.87 \\ 
& \textbf{narrative}    & 0.63   &  NA   & 1.09  & 11.76 \\ 
& \textbf{conversation} & 3.03   & 2.96  & 0.57  & 22.12  \\ 
& \textbf{pet-directed} & 18.75  & 14.57 & 10.00 & NA      \\ 
\bottomrule
 \parbox[t]{4mm}{\multirow{4}{*}{\rotatebox[origin=c]{90}{\parbox[c]{1.5cm}{\centering \textbf{Extrinsic aug.}}}}} &
 \textbf{read}          & 0.98  & 1.89  & 3.37  & 12.50  \\ 
& \textbf{narrative}    & 0.63  & NA    & 1.09  & 11.76  \\ 
& \textbf{conversation} & 4.04  & 2.70  & 1.14  & 18.75  \\ 
& \textbf{pet-directed} & 12.50 & 13.73 & 10.00 & NA      \\ 
\bottomrule
\parbox[t]{4mm}{\multirow{4}{*}{\rotatebox[origin=c]{90}{\parbox[c]{1.5cm}{\centering \textbf{VFR norm.}}}}} &
\textbf{read}           & 0.98  & 1.89  & 3.37  & 18.75  \\ 
& \textbf{narrative}    & 0.48  &  NA   & 1.09  & 11.76  \\ 
& \textbf{conversation} & 3.03  & 2.27  & 1.14  & 18.75  \\ 
& \textbf{pet-directed} & 12.50 & 15.69 & 13.33 & NA     \\ 
\bottomrule
 \parbox[t]{4mm}{\multirow{4}{*}{\rotatebox[origin=c]{90}{\parbox[c]{1.5cm}{\centering \textbf{VFR norm. aug.}}}}} &
 \textbf{read}          & 0.98  & 1.29  & 2.62  & 12.50  \\ 
& \textbf{narrative}    & 0.63  & NA    & 0.55  & 11.76  \\ 
& \textbf{conversation} & 2.69  & 2.27  & 0.38  & 18.75  \\ 
& \textbf{pet-directed} & 12.50 & 12.64 & 14.44 & NA     \\ 
\bottomrule
 \parbox[t]{4mm}{\multirow{4}{*}{\rotatebox[origin=c]{90}{\parbox[c]{1.2cm}{\centering \textbf{Multi-style}}}}} &
  \textbf{read}         & 0.98  & 1.26 & 2.25  & 12.50 \\ 
& \textbf{narrative}    & 0.63  & NA    & 0.73  & 11.76 \\ 
& \textbf{conversation} & 2.02  & 2.27  & 1.14  & 12.50 \\ 
& \textbf{pet-directed} & 12.50 & 15.59 & 13.33 & NA    \\ 
\bottomrule
\end{tabular}
}
\end{table}

In the baseline, a style-mismatch between enrollment and test utterances consistently degraded ASV performance compared to their style-matched task. For instance, when enrolled with conversational speech, the style-matched task (conversation -- conversation) had an EER of 0.57\%. The performance degraded for style-mismatched tasks resulting in EERs of 3.03\%, 2.96\%, and 22.12\% for conversation -- read, conversation -- narrative, and conversation -- pet-directed pairs, respectively. 

The second configuration of extrinsic augmentation performed better than the baseline in 6 tasks out of 14. These tasks were mainly the ones in which the development set was narrative or pet-directed speech. These styles had fewer utterances for adaptation and hence, the increase in the amount of data from augmentation could explain the improvement. On the other hand, the extrinsic augmentation performed worse than the baseline in 3 tasks out of 14 tasks. Interestingly, these were the tasks with reading or conversational speech as the development set. These styles had more utterances than others. 
The standard augmentation techniques used in the \textit{extrinsic augmentation} setup merely increased the amount of data and might not have been sufficient to address style-variability. 


VFR normalization was better than the baseline in 5 tasks out of 14, the same in 4 tasks out of 14, and worse in 5 tasks out of 14. This inconsistency in performance gains between the two setups may be due to: (i) the style normalization from VFR only partially addressed style variability (ii) the VFR normalization was only applied to development data and not to enrollment and test data. We did not apply VFR to enroll and test utterances because it would result in the loss of speaker-specific information. 

The proposed approach of entropy-based VFR normalization augmentation performed better than the baseline in 9 tasks out of 14. The most notable improvement was seen when the testing was on pet-directed speech (read -- pet-directed and conversation -- pet-directed) which is often characterized by exaggerated prosody. However, for two tasks, read -- conversation and pet-directed -- conversation, the proposed approach did not improve the results compared to the baseline. 

When compared to VFR normalization, the proposed approach showed significant improvement in 7 tasks out of 14. The performances were same in 5 tasks out of 14. There was a degradation in performance of the proposed approach for 2 tasks out of 14. 

The proposed approach was better than extrinsic augmentation in 7 tasks out of 14 and the same in 6 tasks out of 14. The proposed approach was generally better even if it used less data than extrinsic augmentation.  This result verifies the hypothesis that VFR, in fact, improved the ASV performance by providing partially style-normalized utterance representations and not by simply increasing the number of samples seen by the PLDA classifier. However, in the pet-directed -- conversation task, characterized by exaggerated prosody, the proposed approach was worse than using extrinsic augmentation. 

The multi-style configuration had more style information available in the development set as compared to the proposed approach--still, their performances were comparable. Their performances were the same in 6 tasks out of 14, 3 tasks out of 14 the proposed was better, and multi-style was better in 5 tasks out of 14. These findings support the hypothesis that VFR methods can be used as a data-augmentation technique when multi-style data are limited. One of the tasks where the proposed approach was better than multi-style was a style-matched task of conversation -- conversation. There are probably variations within a speaking style that could be compensated by the style-normalized augmentation approach. 

\section{Conclusion}
\label{sec:conclusion}
Speaking-style variability degraded ASV performance significantly. The proposed approach used an entropy-based variable frame rate technique to perform data-augmentation when multiple styles were not available to perform an in-domain adaptation of the PLDA classifier. The ASV performance showed significant improvement in the presence of a speaking-style mismatch by partially addressing performance degradation using  VFR data augmentation. The performance of the proposed approach was comparable to the best-case scenario of having multiple styles available for PLDA augmentation. A natural progression of this work is to analyze other possible approaches to address the differences between speaking styles. It would also be interesting to investigate features and/or utterance representation techniques that are less affected by speaking style. More work will need to be done in the future to address the combined effects of speaking-style variability, short duration ($<$ 30~secs), and extrinsic variability on ASV performance.
  
\section{Acknowledgment}
This work was supported in part by the NSF.

\bibliographystyle{IEEEtran}

\bibliography{references}

\end{document}